\documentclass[journal=jctcce,manuscript=article,layout=twocolumn]{achemso}
\usepackage[utf8]{inputenc}
\usepackage[T1]{fontenc}
\usepackage{booktabs}
\usepackage{caption}
\usepackage{threeparttable}
\usepackage{tikz-feynhand}
\usepackage{graphicx}
\usepackage{subcaption}
\usepackage{amssymb}
\usepackage{epstopdf}
\usepackage{amsmath,dsfont}
\usepackage{bm}
\usepackage{commath}
\usepackage{xcolor}
\usepackage{physics}
\usepackage{braket}
\usepackage{hyperref}
\usepackage[capitalize]{cleveref}
\usepackage{float}
\usepackage[justification=raggedright]{caption}
\usepackage{siunitx}
\usepackage{algorithm}
\usepackage{algpseudocode}

\usepackage{stfloats}
\usepackage{placeins}
\usepackage{cuted}

\SectionNumbersOn
\DeclareSIUnit\hartree{\text{\ensuremath{E}}_{\mathrm{h}}}
\DeclareSIUnit\bohr{\text{\ensuremath{a}}_{0}}

\makeatletter

\newcommand{\Rmnum}[1]{\expandafter\@slowromancap\romannumeral #1@}
\makeatother
\newcommand{\E}[1]{\hat{E}_{#1}}
\newcommand{\be}{\begin{equation}}
	
	\newcommand{\ee}{\end{equation}}
\newcommand{\bea}{\begin{eqnarray}}
	\newcommand{\eea}{\end{eqnarray}}
\newcommand{\ba}{\begin{array}}
	\newcommand{\ea}{\end{array}}

\newcommand{\bl}{\begin{flalign}}
	\newcommand{\enl}{\end{flalign}}

\newcommand{\half}{\frac{1}{2}}

\newcommand{\eq}[1]{Eq.~\eqref{#1}}

\newcommand{\fig}[1]{Fig.~\ref{#1}}

\newcommand*{\rom}[1]{\expandafter\@slowromancap\romannumeral #1@}
\renewcommand{\bf}{\mathbf}
\newcommand{\mc}{\mathcal}

\newcommand{\bs}{\begin{split}}
	\newcommand{\es}{\end{split}}

\title{Constrained Optimization Algorithms for Orbital Optimization in Quantum Chemistry}
\author{Junzhe Zhang}
\affiliation{Department of Chemistry and Department of Physics, Westlake University}
\author{Shuoyi Hu}
\affiliation{Department of Chemistry and Department of Physics, Westlake University}
\author{Bing Gu}
\email{gubing@westlake.edu.cn}
\affiliation{Department of Chemistry and Department of Physics, Westlake University}

\begin{document}
	\begin{abstract}

We present a modular constrained-orbital-optimization framework for quantum chemistry. The formulation separates the correlated electronic-structure solver from the orbital optimizer: the solver supplies one- and two-particle reduced density matrices, while the molecular orbitals are updated on the orthonormality-constrained Stiefel manifold with an implicit steepest-descent algorithm. Because the orbital optimizer only requires reduced density matrices, MP2, CASCI, and DMRG can be treated within the same interface. For CASCI solvers, the approach is closely related to optimal-orbital full configuration interaction and CASSCF\cite{helgaker_MulticonfigurationalSelfConsistentField_2000a}, but uses a solver-independent constrained-optimization update rather than CAS-specific orbital-rotation equations. When conventional CASSCF orbital-rotation iterations converge to higher-energy local solutions, CO-CAS can recover lower-energy stationary solutions. We also introduce a modified direct inversion in the iterative subspace procedure to accelerate macro-iteration convergence and a dynamical-weighting scheme to improve state-averaged excited-state calculations. Applications to LiF, H$_2$O, and pyrazine show that orbital optimization lowers energies relative to fixed-orbital MP2, CASCI, and DMRG references while improving convergence and potential-energy-curve smoothness.
		\end{abstract}
	
	\section{Introduction}
	\label{sec:introduction}

A reliable description of ground and excited electronic states is essential for modeling chemical reactivity. For weakly correlated molecules near equilibrium, canonical Hartree--Fock orbitals often provide a useful starting point for post-Hartree--Fock methods. Many chemically important processes, however, are poorly represented in a fixed orbital basis. Bond breaking, photochemical relaxation, and conical intersections can require multiconfigurational wavefunctions, and the exact wavefunction may remain compact only when the orbitals are allowed to adapt to the correlated many-electron state. Orbital optimization provides this variational flexibility by rotating the molecular orbitals while preserving orthonormality. It can lower the total energy, improve the description of correlation, reduce the number of important configurations, and generate smoother potential energy curves. Orbital optimization is therefore not merely a technical refinement of a chosen electronic-structure method; it changes the effective variational space explored within a finite configuration expansion.

The importance of orbital optimization is evident across several electronic-structure methods. In complete active space self-consistent field (CASSCF) theory, the wavefunction is written as a full configuration interaction expansion within a selected active space, and both the configuration coefficients and molecular orbitals are optimized variationally \cite{kreplin_SecondorderMCSCFOptimization_2019a,kreplin_MCSCFOptimizationRevisited_2020a, kudin_BlackboxSelfconsistentField_2002, werner_QuadraticallyConvergentMCSCF_1981, werner_QuadraticallyConvergentMulticonfiguration_1980a}. This simultaneous optimization makes CASSCF a standard method for strongly correlated systems because it treats near-degenerate configurations on an equal footing. Dynamic correlation can then be added perturbatively, for example through CASPT2 \cite{andersson_SecondorderPerturbationTheory_1992, angeli_NelectronValenceState_2001, angeli_QuasidegenerateFormulationSecond_2004, finley_MultistateCASPT2Method_1998b}. Similar motivations appear in other orbital-optimized methods. In orbital-optimized second-order Møller--Plesset perturbation theory (OOMP2), the orbitals are optimized in the presence of perturbative correlation \cite{lochan_OrbitaloptimizedOppositespinScaled_2007a, lee_RegularizedOrbitalOptimizedSecondOrder_2018, bozkaya_QuadraticallyConvergentAlgorithm_2011, bozkaya_OrbitaloptimizedMP25Its_2014, bozkaya_AnalyticEnergyGradients_2013, bozkaya_OrbitalOptimizedMP3MP25_2016, kossmann_CorrelatedInitioSpin_2010}, which can improve MP2 when canonical Hartree--Fock orbitals suffer from orbital instabilities, spin contamination, or symmetry breaking. In density matrix renormalization group self-consistent field (DMRG-SCF), DMRG replaces the active-space full configuration interaction solver in a CASSCF-like optimization framework \cite{zgid_DensityMatrixRenormalization_2008a, ghosh_OrbitalOptimizationDensity_2008}, enabling active spaces much larger than those accessible to conventional CASSCF. More recently, optimal-orbital selected full configuration interaction has formulated orbital optimization from a resource-limited perspective: given a finite number of configurations or orbitals that can be treated exactly, one seeks the orbital subspace that gives the lowest energy under this computational budget \cite{li_OptimalOrbitalSelection_2020a}. These methods share a common structure: the many-electron expansion is limited, and orbital optimization is used to make that finite expansion as expressive as possible.

Orbital optimization can be approached with a variety of algorithms\cite{sun_RecentDevelopmentsPySCF_2020}. In conventional CASSCF implementations, the problem is commonly formulated in terms of orbital rotations between inactive, active, and virtual subspaces. Newton--Raphson \cite{siegbahn_CompleteActiveSpace_1981a} and super-CI algorithms \cite{_CompleteActiveSpace_} have been developed to solve the resulting nonlinear optimization problem and have achieved considerable practical success. These algorithms, however, are often closely tied to the CAS wavefunction form. This dependence is inconvenient when the same orbital-optimization strategy is to be used with different electronic-structure solvers, such as MP2, DMRG, or other correlated methods. A more general perspective is to formulate orbital optimization directly as a constrained optimization problem on the orthonormality manifold. In this formulation, the electronic solver and orbital optimizer can be separated: the solver supplies reduced density matrices (RDMs), while the optimizer updates the molecular orbitals subject to the orthonormality constraint. This separation allows the two components to be developed independently and provides a unified interface for combining orbital optimization with multiple correlated electronic-structure methods.

Motivated by this optimal-orbital perspective, we present a modular constrained-orbital-optimization framework for selected subspaces of the full configuration space. The orbital optimization is carried out with the implicit steepest descent (ISD) algorithm for orthogonality-constrained optimization \cite{oviedo_ImplicitSteepestDescent_2022b}. The complete-active-space limit is included as a special case when $\mc{I} = \mc{I}(n, m)$. Starting from an initial set of molecular orbitals, here chosen as canonical Hartree--Fock orbitals, an electronic-structure solver such as MP2, CASCI\cite{knowles_NewDeterminantbasedFull_1984, olsen_DeterminantBasedConfiguration_1988}, or DMRG \cite{white_DensityMatrixFormulation_1992a,white_DensitymatrixAlgorithmsQuantum_1993a, chan_MatrixProductOperators_2016a, chan_DensityMatrixRenormalization_2011a, wouters_DensityMatrixRenormalization_2014, zhai_Block2ComprehensiveOpen_2023} generates the one- and two-electron RDMs. These RDMs define the orbital-optimization objective, and the molecular orbitals are updated with ISD under the orthonormality constraint. The electronic-structure calculation is then repeated in the updated orbital basis, and the macro-iterations continue until convergence. For a CASCI solver, the resulting CO-CAS method should be interpreted as an alternative orbital-optimization route within the same complete-active-space ansatz; although it does not change the exact variational minimum, it can converge to a lower-energy solution than a conventional CASSCF calculation when the latter is trapped at a higher-energy stationary point. The key point of this work is a modular constrained-orbital-optimization interface that decouples orbital updates from the correlated electronic solver. This allows MP2, CASCI, and DMRG to be treated within the same optimization framework using only the corresponding reduced density matrices.

	We first combine ISD orbital optimization with MP2 (CO-MP2) and apply it to the LiF ground-state potential energy curve (PEC). CO-MP2 lowers the LiF ground-state energy by approximately $0.3$ \si{\milli \hartree} relative to MP2 near equilibrium and gives larger energy lowerings as the bond is stretched. We then combine ISD with CASCI (CO-CAS) and apply it to the ground- and excited-state PECs of LiF. With the 6-31G basis set, CO-CAS and CASSCF give similar ground-state PECs. With the larger aug-cc-pVDZ basis set, CO-CAS gives a smooth and physically reasonable PEC, whereas the CASSCF iterations converge to higher-energy solutions near equilibrium. We next apply CO-CAS to the ground-state energy of H$_2$O at its equilibrium geometry by varying the number of active orbitals while keeping 10 active electrons fixed. With the cc-pVDZ basis set, CO-CAS yields lower energies than CASSCF by approximately 1 \si{\milli \hartree} for all tested active spaces. With the cc-pVQZ basis set, CO-CAS gives slightly higher energies for smaller active spaces, but as the number of active orbitals increases, it again yields energies about 1 \si{\milli \hartree} lower than those from CASSCF. To accelerate macro-iteration convergence, we introduce a modified direct inversion in the iterative subspace (DIIS) procedure \cite{pulay_ConvergenceAccelerationIterative_1980a, pulay_ImprovedSCFConvergence_1982c}, which substantially reduces the number of CO-CAS macro-iterations. For excited states, dynamical weighting gives smoother PECs than state-averaged CO-CAS. Finally, we combine ISD with DMRG (CO-DMRG) for LiF and pyrazine ground-state PECs. CO-DMRG is lower than DMRG by approximately 0.09 \si{\hartree} for LiF and 0.05 \si{\hartree} for pyrazine. All codes are implemented in the open-source program \textsc{PyQED}\cite{xie_PyQEDPythonFramework_2026}.

    The article is organized as follows. In Section \Rmnum{2}, we formulate the constrained-orbital-optimization framework and separate it into two subproblems: the electronic-structure solver and ISD-based orbital optimization. Implementation details, including spin purification, convergence acceleration, and electronic overlap, are discussed in Section \Rmnum{3}. In Section \Rmnum{4}, we apply the method to LiF, H$_2$O, pyrazine ground- and excited-state PECs and analyze convergence. Section \Rmnum{5} summarizes the main conclusions and discusses future extensions.
    
    Atomic units $\hbar = e = m_e = 1$ are used.
	
	\section{Theory}
	We first present the orbital-optimization problem in quantum chemistry. Throughout, $p,q,r,s$ denote generic molecular orbitals (MOs), $i,j,k,l$ denote doubly occupied core orbitals, $t,u,v,w$ denote active orbitals, and $a,b,c,d$ denote unoccupied virtual orbitals. 
		
	\subsection{Orbital Optimization in Quantum Chemistry}

The electronic Hamiltonian in a given set of orthonormal molecular orbitals is given by 
	\be
	H = \sum_{pq}h_{pq} \E{pq} + \half \sum_{pqrs} (pq|rs) \hat{e}_{pqrs}
	\label{eq:h}
	\ee
	where $h_{pq} = \braket{p| -\half \grad^2 + v_\text{eN}(\bf r) | q}$ is the core Hamiltonian and $(pq|rs)$ is the electron-repulsion integral. 
The transition operators are defined as $\E{pq} = \sum_\sigma \hat{c}_{p\sigma}^\dag \hat{c}_{q\sigma}$ with commutation relation $
\sbr{\hat{E}_{pq}, \hat{E}_{rs}} 
= \hat{E}_{ps}\delta_{qr} - \hat{E}_{rq} \delta_{sp}
$     and 	$\hat{e}_{pqrs}  = \E{pq}\E{rs} - \delta_{ps}\E{qr}$ \cite{belcher_GradientsNonAdiabaticDerivative_2011}.
%
%
%

	It is interesting to note that the transition operators defined by $\hat{X}_{PQ} = \ket{P}\bra{Q}$, where $\ket{P}$ denotes an orthonormal many-body state of a system, satisfy the same commutation relation as the excitation operators. This suggests that there is a one-to-one mapping between an arbitrary Hilbert space of size $N$ and an electronic Fock space, subject to a constraint on the electron number $\hat{N} = 1$ \cite{gu_CollectiveChemistryStrong_2025}.  Such mapping can be used to transform a generic Hamiltonian to a coupled fermion-boson problem that can be tackled by many-body perturbation theory \cite{gu_CollectiveChemistryStrong_2025}.

For wavefunction-based multireference quantum chemistry methods, such as complete active space configuration interaction (CASCI), the many-electron wavefunction is expanded as a linear combination of configurations,
	\be \ket{ \Psi}_\mc{I} = \sum_{I \in \mc{I}} c_I \ket{\Phi_I} 
	\ee 
	The choice of subspace, labeled by the set of binary strings $\mc{I}$, defines the CI method. 
	In the CASCI approach, the index set includes all configurations obtained by
	distributing \(n\) electrons among \(m\) preselected active orbitals. Abelian symmetries, such as electron number and total spin projection \(S_z\), can be built into the index set. The electronic energy with orbital optimization is defined as
	\be 
	E = \min_{\bf U, \bf c} E[\bf U, \bf c; \mc{I}].
	\ee 
	where $\bf U$ denotes the orbital-rotation matrix and $\bf c$ denotes the configuration coefficients.
	For fixed configuration coefficients, the orbital-dependent energy is
\begin{equation}
\begin{aligned}
	E[\bf U; \bf c] = \sum_{pq} h_{pq}\gamma_{pq} U_{pp'} U_{qq'}  \\
    + \half \sum_{pqrs} (pq|rs) \Gamma_{pqrs} U_{pp'} U_{qq'} U_{rr'} U_{ss'}
\end{aligned}
\end{equation} 
	where \(U\) is the orbital-rotation matrix. 
		\be \gamma_{pq} = \braket{\Psi| \E{pq} | \Psi} 
		\label{eq:rdm1}\ee
		is the single-electron reduced density matrix (1-RDM) and 
	
	\be \Gamma_{pqrs} = \braket{\Psi| \hat{e}_{pqrs} | \Psi} 
	\label{eq:rdm2}\ee
	is the two-electron reduced density matrix (2-RDM). 

	The stationary conditions are
	\be
	\begin{split}
	\frac{\partial E[\bf U, \bf c]}{\partial \bf U} &= 0  \\  
	\frac{\partial E[\bf U, \bf c]}{\partial \bf c} &= 0 
		\end{split}
	\label{eq:sta_con}
	\ee 
	The same idea is used for the other electronic-structure solvers discussed in Sections~\ref{sec:mp2} and \ref{sec:dmrg}.

\subsection{Constrained Orbital Optimization with Implicit Steepest Descent}
\label{sec:orbopt}

Orbital optimization can be viewed as an optimization problem with orthonormality constraints. Numerical algorithms for such constrained problems have advanced substantially in applied mathematics \cite{jiang_FrameworkConstraintPreserving_2015, wen_FeasibleMethodOptimization_2013a, _RiemannianBFGSMethod_, absil_OptimizationAlgorithmsMatrix_2008, nocedal_ConjugateGradientMethods_1999}, making them promising tools for quantum chemistry applications. 

Here we implement the implicit steepest descent algorithm of Ref. \cite{oviedo_ImplicitSteepestDescent_2022b} for the orbital optimization.
%
%
%
%
We briefly describe the ISD algorithm here and provide implementation details in Section~\ref{sec:imp isd}.
The optimization problem minimizes the electronic energy $P_4(\bf U) \equiv E[\bf U; \bf c]$, a fourth-order polynomial of the rotation matrix $U$ 
\be
\mc{L}(U, \Lambda) = P_4(U) - \half \Tr\qty[   \Lambda^\top \qty( U^\top U - I_p) ] 
\ee 
with the orthonormality constraint \(U^{\top} U=I\), where \(\Lambda\) is the Lagrange multiplier. Differentiating the Lagrangian with respect to \(U\) and \(\Lambda\) gives
\begin{subequations}
	\label{eq:eq 1st order condition}
	\begin{align}
		\nabla_U \mc{L}(U, \Lambda) &= G-U\Lambda =0,
		\label{eq:eq_1st_order_condition_a}\\
		\nabla_\Lambda \mc{L}(U, \Lambda) &= U^\top U - I_p = 0.
		\label{eq:eq_1st_order_condition_b}
	\end{align}
\end{subequations}
where $G=\nabla_U P_4(U)$ is the Euclidean gradient. Then $U$ satisfies \eq{eq:eq 1st order condition} with Lagrangian multiplier $\Lambda=G^\top U$. Defining
\begin{equation}
	\nabla P_4(U) \equiv G - UG^\top U \text{ and } A(U) \equiv GU^\top - UG^\top
\end{equation}
then $\nabla P_4(U) = A(U)U$, is a constraint-preserving gradient-like direction, the update generated by this direction is compatible with the orthonormality constraint $ U^\top U = I_p$.


Standard gradient descent such as backward Euler method can iteratively update $U_k$ on the manifold $\mc{M}$  
\begin{equation}
	U_{k+1} = U_k - \tau_k \nabla_\mc{M} P_4(U_{k+1})
	\label{eq: euler method}
\end{equation}
Since the iterative process \eq{eq: euler method} is nonlinear, it is difficult to implement this approach. To simplify this problem, $\nabla_\mc{M} P_4(U_{k+1})$ can be approximated by
\begin{equation}
	\nabla_\mc{M} P_4(U_{k+1}) = A(U_{k+1})U_{k+1} \approx A(U_k)U_{k+1} 
	\label{eq:appro}
\end{equation}
Combining \eq{eq: euler method} and \eq{eq:appro}, yields the implicit gradient scheme
\begin{equation}
	\begin{split}
		&U_{k+1}(\tau) = (I_n + \tau_k A(U_k))^{-1} U_k
	\end{split}
\end{equation}
where $\tau$ is the step size. Following Oviedo et al.~\cite{oviedo_ImplicitSteepestDescent_2022b}, the Barzilai-Borwein step is used because it is known to greatly speed up the convergence of gradient-type methods \cite{barzilai_TwoPointStepSize_1988}.
Since $U_{k+1}$ may not satisfy the orthonormal condition in \eq{eq:eq 1st order condition}, it is projected back onto the manifold, defined as
\be
\pi(U) = \arg \min_{X \in \text{St}(n,p)} \norm{U - X}_\text{F} 
\label{eq: PI(U)}
\ee 
where $\norm{X}_\text{F} = \sqrt{\Tr{X^\top X}}$ is the Frobenius norm. \eq{eq: PI(U)} finds the matrix in the Stiefel manifold closest to the given matrix under the Frobenius norm. 
For $p \le n$, the projection operator can be computed by  the eigenvalue decomposition of  $U^\top U = VDV^\top$,  where $D$ is a diagonal matrix containing the eigenvalues, 
\be \pi(U) = UVD^{-1/2}V^\top \ee
So given the $k$th-step point $U_k \in \texttt{St}(n, p)$ (the MO coefficients are $C_k = C_0 U_k$),  the update  is determined by 
\be
U_{k+1}(\tau) = \pi\qty( (I_n + \tau_k A(U_k))^{-1} U_k  ) 
\ee 


%
%
%

\subsection{Electronic Structure Solver}
We consider three methods, including MP2, CAS and DMRG.

\subsubsection{MP2}
\label{sec:mp2}
For the second order Møller–Plesset perturbation theory (MP2), the total Hamiltonian is written as\cite{_ModernQuantumChemistry_}
\begin{equation}
	\hat{H}=\hat{H}^{(0)}+\hat{V}
\end{equation}
where $\hat{H}^{(0)}$ is zeroth-order Hamiltonian and $\hat{V}$ is perturbation term.
It uses the Hartree-Fock (HF) calculation as the starting point\cite{helgaker_HartreeFockTheory_2000}, the zeroth-order Hamiltonian $\hat{H}^{(0)}$ is the summation of Fock operators
\begin{equation}
	\hat{H}^{(0)}=\sum_\mathbf{r} \hat{F}(\mathbf{r}) = \sum_\mathbf{r} \hat{h}(\mathbf{r}) + \sum_{\mathbf{r},i} \left[\hat{J}_i(\mathbf{r})+\hat{K}_i(\mathbf{r})\right]
\end{equation}
where $\hat{h}(r)$ is the one-electron kinetic and electron--nuclear attraction operator. The Coulomb and exchange operators for electron  in orbital $\chi_i$ are defined by
\begin{align}
	\hat{J}_i(\mathbf{r}_1)\chi_j(\mathbf{r}_1)
	&=
	\left[\int \dd{\mathbf{r}_2}\,
	\chi_i^*(\mathbf{r}_2)\frac{1}{\abs{\mathbf{r}_{12}}}\chi_i(\mathbf{r}_2)\right]
	\chi_j(\mathbf{r}_1),\\
	\hat{K}_i(\mathbf{r}_1)\chi_j(\mathbf{r}_1)
	&=
	\left[\int \dd{\mathbf{r}_2}\,
	\chi_i^*(\mathbf{r}_2)\frac{1}{\abs{\mathbf{r}_{12}}}\chi_j(\mathbf{r}_2)\right]
	\chi_i(\mathbf{r}_1).
\end{align}
%

The MP2 correction energy is
\begin{equation}
	E^{(2)}_{\text{MP}}=\sum_{\mu>0} \frac{\abs{V_{0\mu}}^2}{E_0-E_{\mu}} =\frac{1}{4}\sum_{ij}^{\text{occ}}\sum_{ab}^{\text{vir}} \frac{\abs{(ia||jb)}^2}{\epsilon_i+\epsilon_j-\epsilon_a-\epsilon_b}
\end{equation}
where $(ia||jb)=(ia|jb) - (ij|ab)$. By Brillouin's theorem, \(\braket{\Phi_0|\hat{V}|\Phi_{i}^a}=0\), so only doubly excited configurations \(\Phi_\mu=\Phi_{ij}^{ab}\) contribute to \(V_{0\mu}\) through the Slater--Condon rules. The MP2 amplitudes generate the 1- and 2-RDMs; detailed derivations are provided in the Supporting Information, Section S1. For MP2 with constrained orbital optimization (CO-MP2), the configuration coefficient is fixed by the single-reference configuration, so only \eq{eq:eq_1st_order_condition_a} needs to be considered.

CO-MP2 is algorithmically different from fully variational OOMP2. In OOMP2, the MP2 Lagrangian or Hylleraas functional is optimized directly with respect to orbital rotations, using explicit orbital gradients and, in Newton-type implementations, orbital Hessians \cite{lochan_OrbitaloptimizedOppositespinScaled_2007a, lee_RegularizedOrbitalOptimizedSecondOrder_2018, bozkaya_QuadraticallyConvergentAlgorithm_2011}. In contrast, CO-MP2 keeps the MP2 RDMs fixed during each orbital-update step and rebuilds the MP2 amplitudes only after the macro-iteration.

\subsubsection{CASCI}
CASCI partitions the molecular orbitals into doubly occupied core orbitals, active orbitals, and unoccupied virtual orbitals. The CASCI wavefunction is built from the Slater determinants obtained by distributing all active electrons among the preselected active orbitals.
Thus, the CASCI wavefunction is a linear combination of Slater determinants,
	$\ket{\Psi} = \sum_I c_I \ket{I}$.
For a given set of orbitals and CI coefficients, the total energy is 
\begin{equation}
	E_{\text{tot}} = \sum_{tu} F_{tu} \gamma_{tu} + \frac{1}{2} \sum_{tuvw} (tu|vw) \Gamma_{tuvw} + E_{\text{core}}
\end{equation}
where the closed-shell Fock matrix \(F_{tu}\) is defined as \(F_{tu} = h_{tu} + \sum_i \left[2(tu|ii) - (ti|iu)\right]\), and \(E_\text{core}\) is the core-electron energy.
 
\(\gamma_{tu}\) and \(\Gamma_{tuvw}\) are the active-space 1- and 2-RDMs. For fixed orbitals, they depend only on the CI coefficients and determinant basis. For a CASCI wavefunction, the RDMs in \eq{eq:rdm1} and \eq{eq:rdm2} read
\be \gamma_{pq} = \sum_{I, J \in \mc{I}} c_I^*c_J\braket{I| \E{pq} | J} \ee
\be \Gamma_{pqrs} =  \sum_{I, J \in \mc{I}} c_I^*c_J\braket{I| \hat{e}_{pqrs} | J} \ee
These RDMs are the inputs for constrained orbital optimization in Section \ref{sec:orbopt}. The quantities $\gamma_{tu}$ and $\Gamma_{tuvw}$ are derived in the Supporting Information, Section S2.

 CO-CAS is closely related to CASSCF because both methods optimize orbitals in a complete active space. The difference lies in the orbital-optimization strategy: conventional CASSCF solves coupled CI--orbital stationarity conditions with CAS-specific orbital-rotation algorithms\cite{kreplin_SecondorderMCSCFOptimization_2019a,kreplin_MCSCFOptimizationRevisited_2020a}, whereas CO-CAS follows a general fixed-RDM procedure in which the CASCI RDMs are held fixed during each constrained ISD orbital update and regenerated in the next macro-iteration. Thus, CO-CAS can be viewed as a solver-independent constrained-optimization alternative to conventional CASSCF.

\subsubsection{DMRG}
\label{sec:dmrg}
CASCI scales exponentially and therefore becomes infeasible for large active spaces.
In contrast to CASCI, which performs exact diagonalization over all active-space configurations, DMRG represents the many-electron wavefunction with a compressed matrix product state (MPS) ansatz \cite{white_DensityMatrixFormulation_1992a, white_DensitymatrixAlgorithmsQuantum_1993a, schollwock_DensitymatrixRenormalizationGroup_2011b,chan_DensityMatrixRenormalization_2011a, chan_AlgorithmLargeScale_2004, chan_HighlyCorrelatedCalculations_2002}.

\be \ket{\Psi} = \sum_{a_i,n_i} A^{n_1}_{1a_1} A^{n_2}_{a_1a_2} ... A^{n_L}_{a_{n-1}1} \ket{n_1 n_2 ... n_L}
\label{eq:MPS_ansatz}
\ee
where \(L\) is the number of sites and \(\ket{n_1 n_2 ... n_L}\) is an occupation-number vector. Each local basis state \(\ket{n_i}\) is a spin-orbital occupation with two possibilities, \(\ket{0}\) and \(\ket{1}\). The tensor \(A^{n_i}_{\alpha_{i-1}\alpha_i}\) has one physical index \(n_i\) and two virtual indices \(\alpha_{i-1}\) and \(\alpha_i\), whose dimensions are bounded by the bond dimension \(D\). The MPS wavefunction becomes equivalent to full CI in the \(D \rightarrow \infty\) limit. By analogy with \eq{eq:MPS_ansatz}, the Hamiltonian can be expressed as a matrix product operator (MPO) \cite{ren_GeneralAutomaticMethod_2020a}
\begin{equation}
\begin{aligned}
	\mc{W} = \sum_{n_i,n_i',b_i} W_{1,b_1}^{n_1,n_1'} W_{b_1, b_2}^{n_2, n_2'} ... W_{b_{L-1},1}^{n_L, n'_L} \\
    \ket{n_1 n_2 ... n_L} \bra{n_1' n_2' ... n_L'}
	\label{eq:MPO_ansatz}
\end{aligned}
\end{equation}
where each MPO tensor $W^{n_i,n_i'}_{b_{i-1},b_i}$ has two physical indices $n_i, n_i'$ and two virtual indices $b_{i-1}, b_i$.

Combining \eq{eq:MPS_ansatz} and \eq{eq:MPO_ansatz}, finding the ground state is equivalent to minimizing the  Lagrangian function $\mc{L} = \braket{\Psi|\hat{H}|\Psi} - \lambda\left(\braket{\Psi|\Psi}-1\right)$, with the stationary condition written with MPS/MPO language
\begin{equation}
	\begin{split}
		& \frac{\partial \mc{L}}{\partial A_{a_{i-1},a_i}^{n_i \ast}} = \sum_{n_i'}\sum_{a_{i-1}'a_i'b_{i-1}b_i}L_{b_{i-1}}^{a_{i-1}a_{i-1}'} W_{b_{i-1},b_i}^{n_{i},n_i'} R_{b_{i}}^{a_{i}a_{i}'} A_{a'_{i-1},a'_i}^{n'_i} \\
        &- \lambda \sum_{a_{i-1}' a_i'} \Psi_{a_{i-1}a_{i-1}'}^L \Psi_{a_i a_i'}^R A_{a_{i-1}' a_i'}^{n_i} = 0 \\
		& \frac{\partial \mc{L}}{\partial \lambda} = \sum_{a_i,a_i'} \left(\sum_{n_1}A_{1,a_1}^{n_1 \ast} A_{1,a'_1}^{n_1}\right)	\times ... \times \left(\sum_{n_i}A_{a_{i-1},a_i}^{n_i \ast} A_{a'_{i-1},a'_i}^{n_i}\right) \\
		&\times ... \times \left(\sum_{n_L}A_{a_{L-1},1}^{n_L \ast} A_{a'_{L-1},1}^{n_L}\right) = 0
	\end{split}
\end{equation}
where
\begin{equation}
	\begin{split}
		& L_{b_{i-1}}^{a_{i-1}a_{i-1}'} = \sum_{a_i,b_i,a_i'}\left(\sum_{n_1, n_1'}A_{1,a_1}^{n_1 \ast} W_{1,b_1}^{n_1,n_1'}A_{1,a'_1}^{n'_1}\right) \times ... \\
        &\times \left(\sum_{n_{i-1}, n_{i-1}'}A_{a_{i-2},a_{i-1}}^{n_{i-1} \ast} W_{b_{i-2},b_{i-1}}^{n_{i-1},n_{i-1}'}A_{a'_{i-2},a'_{i-1}}^{n'_{i-1}}\right) \\
		& R_{b_{i}}^{a_{i}a_{i}'} =  \sum_{a_i,b_i,a_i'}  \left(\sum_{n_{i+1}, n_{i+1}'}A_{a_{i},a_{i+1}}^{n_{i+1} \ast} W_{b_{i},b_{i+1}}^{n_{i+1},n_{i+1}'}A_{a'_{i},a'_{i+1}}^{n'_{i+1}}\right) \\
        &\times ...  \times \left(\sum_{n_{L}, n_{L}'}A_{a_{L-1},1}^{n_L \ast} W_{b_{L-1},1}^{n_{L},n_L'}A_{a'_{L-1},1}^{n'_L}\right) \\	
		& \Psi_{a_{i-1}a_{i-1}'}^L = \sum_{n_i}\left(A^{n_{i-1}\dagger} ... A^{n_{1}\dagger}A^{n_{1}} ... A^{n_{i-1}}\right)_{a_{i-1},a_{i-1}'} \\
        &\Psi_{a_i a_i'}^R = \sum_{n_i} \left(A^{n_{i+1}} ... A^{n_L}A^{n_L\dagger} ... A^{n_{i+1}\dagger}\right)_{a_i',a_i}
	\end{split}
\end{equation}

 This formulation allows the ground-state wavefunction to be solved variationally
using a sweeping algorithm that sequentially optimizes two local tensors $A_{n_i}A_{n_{i+1}}$ while keeping all other environmental tensors fixed. Effectively, this means that we solve a sequence of eigenvalue problems in reduced Hilbert spaces, whose final convergence accuracy is controlled by the bond dimension $D$. In our implement, we also apply abelian symmetry to MPS during the sweep, keeping the sector with target quantum number i.e. particle number and $S_z$.

To combine DMRG with constrained orbital optimization, the 1- and 2-RDMs are required as inputs. In constrained-optimization DMRG (CO-DMRG), the MPS tensors \(A^{n_i}\) and molecular orbitals are optimized in alternating macro-iterations, analogous to the CAS stationarity conditions in \eq{eq:sta_con}. The one- and two-particle RDMs can be expressed as tensor contractions over the physical indices,
\be
\begin{split}
\gamma_{pq} = &\sum_{\{n_i\},\{n'_j\},\sigma} \big[A^{n_1} ... A^{n_i} ... A^{n_L}\big]^\dagger A^{n'_1} ... A^{n'_j} ... A_{n'^L} \\
&\braket{n_1...n_i ...n_L | c^{\dagger}_{p\sigma} c_{q\sigma} |n'_1...n'_j...n'_L}
\end{split}
\ee
and similarly for $\Gamma_{pqrs}$.   A detailed derivation of the DMRG RDMs is provided in the Supporting Information, Section S3.

CO-DMRG and DMRG-SCF both combine DMRG with orbital optimization and use DMRG 1- and 2-RDMs to update the orbitals, but they differ in the optimization strategy. In DMRG-SCF, the orbital rotations are usually formulated in close analogy to CASSCF: the DMRG solver replaces the FCI solver in the active space, and the orbitals are optimized self-consistently with respect to the DMRG active-space wave function\cite{zgid_DensityMatrixRenormalization_2008a}. In contrast, CO-DMRG follows a solver-independent fixed-RDM strategy in which the DMRG RDMs are held fixed during each constrained ISD orbital update and the DMRG wavefunction is recomputed in the next macro-iteration.
%

\subsubsection{Excited States with State-Averaged and Dynamical-Weighting }
For excited-state calculations, state averaging (SA) changes the orbital-optimization objective to the weighted average energy
\be
E_\text{SA} = \sum_\alpha w_\alpha E_\alpha[U] 
\ee 
where $w_\alpha > 0$ are the weights satisfying $\sum_\alpha w_\alpha = 1$. This simply amounts to the state-averaged reduced density matrices $\bar{\gamma} = \sum_\alpha w_\alpha \gamma_\alpha$ and $\bar{\Gamma} = \sum_\alpha w_\alpha \Gamma_\alpha$ .

However, we found that the state-averaged (SA) method is not suitable for CO-CAS because potential energy curves (PECs) in some bond-stretching problems given by SA-CO-CAS can become discontinuous, for example in the LiF bond-stretching calculation discussed below. To overcome this difficulty, we introduce a dynamical weighting (DW) scheme that can automatically change the weights during the CO-CAS calculation and gives smooth PECs.  The DW-CO-CAS energy is written as a weighted summation:

\begin{equation}
	E^{\text{DW}} = \sum_{i}^{N} \omega_i(E^{\text{CAS}}) E^{\text{CAS},i}
	\label{dw_energy}
\end{equation}
where $E^{\text{CAS}}$ is the $i$th state CAS energy and $\omega_i$ is the $i$th state weight, which is related to CAS energy and sum over all $N$ states included in the SA calculation. The expressions of weight $\omega_i$ are


\begin{equation}
\omega_i =
\begin{cases}
1/W, & i \leq L,\\
\begin{aligned}
&\bigl[1-3(\Delta E_{i,I}/\alpha)^2 \\
& +2(\Delta E_{i,I}/\alpha)^3\bigr]/W,
\end{aligned}
& i>L,\ \Delta E_{i,I}<\alpha,\\
0, & i>L,\ \Delta E_{i,I}\geq\alpha .
\end{cases}
\end{equation}

where the weight are set equal for the L lowest energy states and tapered to zero with a polynomial spline for the higher energy states, $\Delta E_{i,I} = E^{\text{CAS}, i} - E^{\text{CAS}, I}$, $E^{\text{CAS}, I}$ is the energy of a state of interest and $\alpha$ is the energy range of the spline taper.

	\section{Computational Methods}
	\label{sec:implement}
	In this section, we describe the constrained-orbital-optimization algorithm and discuss implementation details, including spin purification, convergence acceleration, and electronic overlap.


\subsection{Constrained Orbital Optimization with ISD}
\label{sec:imp isd}
This section introduces the constrained orbital-optimization algorithm based on the implicit steepest descent (ISD) method.
The implicit steepest descent (ISD) optimization procedure for minimizing
$P_4(U)$ under the orthogonality constraint $U^\top U=I_p$ can be summarized
as follows.

\begin{enumerate}
	\item
	Choose an initial guess \(U_0\in \mathrm{St}(n,p)\), an initial step size \(\tau>0\), lower and upper bounds \(0<\tau_m\leq \tau_M\), and parameters \(\eta\in[0,1)\), \(\rho_1,\epsilon,\delta\in(0,1)\). Set \(Q_0=1\), \(C_0=P_4(U_0)\), and \(k=0\).
	
	\item
	At the current iterate \(U_k\), compute $\|\nabla P_4(U_k)\|_F$, where $\|A\|_F = \sqrt{\operatorname{Tr}[A^TA]}$ is the Frobenius norm. If $\|\nabla P_4(U_k)\|_F \leq \epsilon$, stop the algorithm.
	
	
	\item 
	Update the step size by \(\tau=\delta\tau\) until the nonmonotone line-search condition
	\begin{equation}
		P_4(Y(\tau)) \leq C_k + \rho_1 \tau
		\frac{\partial P_4[U_k]}{\partial U_k}[\dot{Y}(0)]
	\end{equation}
	is satisfied, where $\dot{Y}(0)=-\nabla P_4(U_k)$ and $Y(\tau)=\pi\left(\left(I_n+\tau A(U_k)\right)^{-1}U_k\right)$.
	
	\item
	Update $U_{k+1} = Y(\tau)$.
	
	\item 
	Calculate $Q_{k+1} = \eta Q_k + 1$ and $C_{k+1} = \dfrac{\eta Q_k C_k + P_4(U_{k+1})}{Q_{k+1}}$.
	
	\item 
	Use Barzilai-Borwein step sizes (BB steps) to determine $\tau = \tau_{k+1}^{\text{ABB}}$. The definition of BB steps is
	\begin{equation}
		\tau_k^{ABB} = \left\{
		\begin{aligned}
			&\frac{\|S_{k-1}\|_F^2}{\left|\langle S_{k-1},\, Y_{k-1}\rangle\right|}&, \quad &\text{for odd }k;\\
			&\frac{\left|\langle S_{k-1},\, Y_{k-1}\rangle\right|}{\|Y_{k-1}\|_F^2}&, \quad &\text{for even }k.
		\end{aligned}
		\right.		
	\end{equation}
	where $S_{k-1}=U_k-U_{k-1}$ and
	$Y_{k-1}=\nabla P_4(U_k) - \nabla P_4(U_{k-1})$.
	Set
	\[
		\tau=\max\{\min\{\tau_{k+1}^{\mathrm{ABB}},\tau_M\},\tau_m\}.
	\]
	
	\item
	Set $k = k+1$, repeat steps $2-7$.
\end{enumerate}
	
%
%
%

\subsection{Overall Workflow}
The overall constrained orbital-optimization algorithm is as follows:
\begin{enumerate}
	\item  Set the iteration index $k=0$, perfrom a HF calculation to obtain the canonical molecular orbitals (MOs) $C_0$. 
	\item  Calculate one-electron and two-electron integrals with MOs $C_k$.
	\item  perform electronic structure calculation (MP2, CASCI, DMRG), get electronic energy, CI coefficients $c_k$ and calculate 1-RDMs, 2-RDMs with $c_k$. 
	\item If the decay of energy is lower than given tolerance, the iterations converged and algorithm stopped.
	\item Optimize the orbitals by the ISD algorithm, generate the orbital rotation matrix $U_k$, MOs are updated as $C_{k+1}=C_0U_k$.
    \item  Set $k = k+1$ and repeat steps $2-6$.
	\end{enumerate}

	In the following, we discuss some details of the algorithm, i.e., spin purification, convergence acceleration with direct inversion in the iterative subspace (DIIS) and electronic overlap.
	
	\paragraph*{Spin Purification} For Slater determinant-based implementation, the number of electrons and $z$-component of the total spin $\hat{S}_z$ quantum number can be easily fixed. Nevertheless, the eigenstates are not necessarily eigenstates of the total spin operator $ \hat{\bf S}^2$ with $\hat{\bf S} = \sum_i \hat{\bf S}_i$. 
	The total spin can be purified by adding an energy penalty to the Hamiltonian\cite{weser_SpinPurificationFullCI_2022}.  For singlet states, we modify the Hamiltonian by 
	\be
	\hat{H}_\text{SP} =  \hat{H} + \mu \hat{\bf S}^2 
	\ee 
	where the energy panelty term with $\mu > 0$ does not change the singlets $S = 0$ but increases the energy of $S \ne 0$.  For non-singlet states, one possibility is 
	\be
	\hat{H}_\text{SP} = \hat{H} + \mu (\hat{\bf S}^2 - S(S+1))^2 
	\ee 
	
	The total spin operator is
	\be
	\hat{\bf S}^2 = \sum_i \hat{S}_i^2 + \sum_{i \ne j} \hat{S}_i \cdot \hat{S}_j
	\ee
	and the spin-squared operator is
	\be
	\hat{S}_i^2 = \frac{3}{4} (\hat{E}_{ii} - \hat{e}_{ii,ii})
	\ee
	and the spin-spin interaction is
	\be
	\hat{S}_i \cdot \hat{S}_j = -\frac{1}{2} (\hat{e}_{ij,ji} + \frac{1}{2} \hat{e}_{ii, jj}), j \ne i
	\ee 
	
	so $ \hat{\bf S}^2 = \sum_p \frac{3}{4} \E{pp} - \sum_{p,q}  \half \qty( \E{pq,qp} + \half \E{pp, qq}  ) $.
	
	\paragraph*{Convergence Acceleration}
	We introduce a modified DIIS method to speed the convergence of macro-iterations. We store a set of orbital rotation matrices $\{U_k\}$ generated from the orbital optimization procedure at macro-iteration $k$. A set of residual matrices defined as
	\begin{equation}
		\Delta U^k = U^{k+1} - U^k
	\end{equation}
	DIIS assumes that the final solution $U^f$ can be approximated as a linear combination of previous guess matrices
	\begin{equation}
		U=\sum_{i=1}^{m} c_iU^i
		\label{eq:approx}
	\end{equation}
	where $m$ is the number of matrices we choose to approximate the final solution. And $c_i$ are obtained by minimizing the associated residual matrices in a least-square case $\braket{\Delta U|\Delta U}$
	\begin{equation}
		\Delta U = \sum_{i=1}^{m} c_i|\Delta U^i|
	\end{equation}
	with the normalization condition $\sum_{i=1}^{m} c_i=1$.
	
	These requests can be satisfied by minimizing the following Lagrangian function
	\begin{equation}
		\mc{L}=c^{\dagger}Bc-\lambda\left(1-\sum_{i=1}^{m} c_i\right)
	\end{equation}
	where $c$ is coefficients of \eq{eq:approx}, $B$ is the overlap matrix $B_{ij}=\braket{\Delta U^i|\Delta U^j}$.The stationary conditions are
	\begin{equation}
		\begin{split}
			& \frac{\partial \mc{L}}{\partial c_k} = 2\sum_{i=1}^{m} c_iB_{ki} - \lambda = 0 \\
			& \frac{\partial \mc{L}}{\partial \lambda} = 1-\sum_{i=1}^{m} c_i = 0
		\end{split}
		\label{eq:diis}
	\end{equation}
	\eq{eq:diis} can be rewritten as a matrix equation
		\[
	\begin{pmatrix}
		B_{11} & \cdots & B_{1m} & -1 \\
		\vdots & \ddots & \vdots & \vdots \\
		B_{m1} & \cdots & B_{mm} & -1 \\
		-1     & \cdots & -1     & 0
	\end{pmatrix}
	\begin{pmatrix}
		c_1 \\ \vdots \\ c_m \\ \lambda
	\end{pmatrix}
	=
	\begin{pmatrix}
		0 \\ \vdots \\ 0 \\ -1
	\end{pmatrix}
	\]
	Because a linear combination of $U = \sum_{i=1}^{m} c_i U^i$ is not guaranteed to remain unitary isometric,
	we re-orthonormalizes it with polar decomposition
	\begin{equation}
		U_{\mathrm{orth}} = U \left(U^\dagger U\right)^{-1/2}
	\end{equation}


\section{Results and Discussion}
We demonstrate the performance of the constrained-orbital-optimization framework with tests designed to examine three complementary aspects of the method. First, CO-MP2 is used to show that orbital optimization can improve a perturbative correlation treatment along the LiF bond-stretching coordinate relative to conventional MP2.
	 Second, CO-CAS is compared with conventional CASSCF to validate the constrained-optimization formulation in a complete active space and to examine the role of DIIS acceleration. Third, CO-DMRG is used to demonstrate that ISD orbital optimization can be coupled to a tensor-network solver and applied to both LiF and pyrazine.

For all potential energy curves (PECs), the same basis set and active space are used for the optimized and unoptimized calculations in each comparison, so that the energy differences directly reflect the effect of orbital optimization.

\subsection{CO-MP2: Orbital Optimization Improves MP2 for Stretched LiF}
\begin{figure}[htbp]
	\centering
	\includegraphics[width=0.5\textwidth]{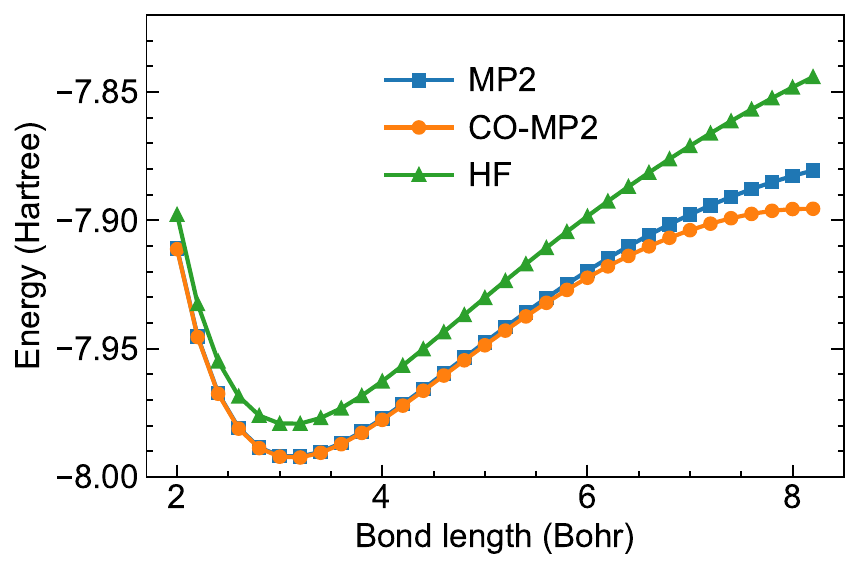} 
	\caption{LiF ground-state potential energy curves calculated with CO-MP2 and MP2 using the 6-31G basis set.}
	\label{fig:mp2}
\end{figure}

We first use CO-MP2 to calculate the LiF ground-state PEC with the 6-31G basis set, as shown in \fig{fig:mp2}. Along the LiF stretching coordinate, the correlation energy increases from 0.013 \si{\hartree} at 2.2 \si{\bohr} to 0.036 \si{\hartree} at 8.2 \si{\bohr}, indicating that correlation effects become increasingly important as the bond is stretched. 

CO-MP2 and MP2 give PECs with similar shapes, but CO-MP2 consistently lowers the energy. Near equilibrium, the improvement is modest: CO-MP2 is lower than MP2 by approximately $3\times10^{-4}$ \si{\hartree}, as expected because canonical HF orbitals already provide a reasonable zeroth-order reference in this region. In the bond-stretching region, the lowering becomes larger and reaches 0.015 \si{\hartree} at 8.2 \si{\bohr}. This behavior shows that as the bond is stretched and static correlation becomes more important, the single-reference HF orbital basis becomes less optimal. These results indicate that constrained orbital optimization does not qualitatively change the MP2 PEC in the weakly correlated region, but it systematically improves the energy as static-correlation increases.

\subsection{CO-CAS: CASSCF Benchmark, DIIS Acceleration, and Excited States}
\begin{figure}[htbp]
	\centering
	\begin{subfigure}[b]{0.45\textwidth}
		\centering
		\includegraphics[width=\textwidth]{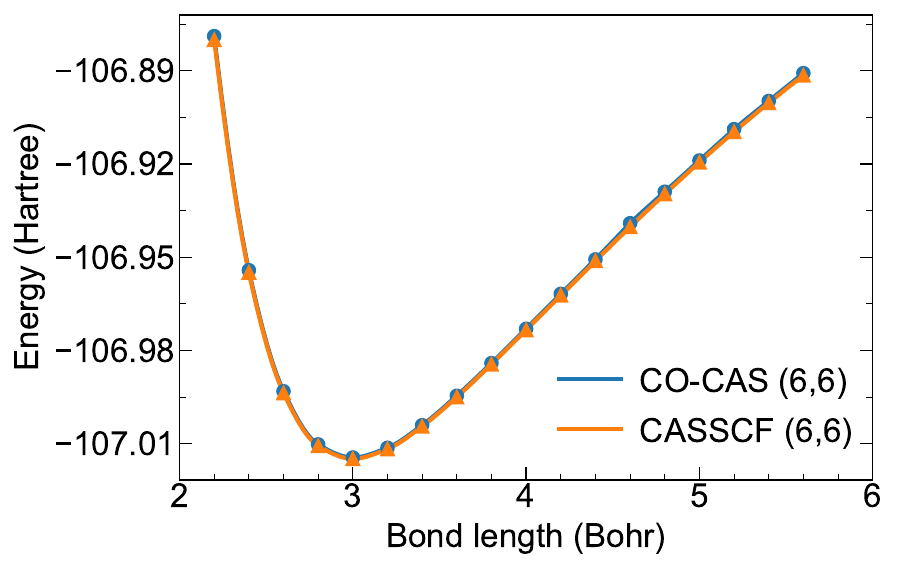}
		\caption{6-31G basis set}
		\label{fig:cas_631g}
	\end{subfigure}
	\hfill 
	\begin{subfigure}[b]{0.45\textwidth}
		\centering
		\includegraphics[width=\textwidth]{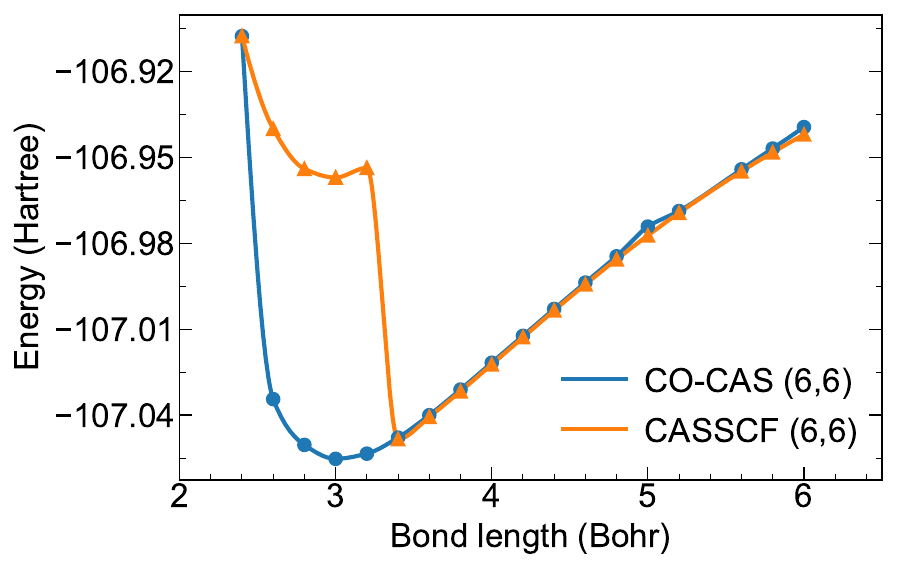}
		\caption{aug-cc-pVDZ basis set}
		\label{fig:cas_augccpvdz}
	\end{subfigure}
	
	\caption{LiF ground-state PECs calculated with CO-CAS(6,6) and CASSCF(6,6) using the 6-31G and aug-cc-pVDZ basis sets.}
	\label{fig:two_images_1}
\end{figure}

We next combine constrained orbital optimization with a CASCI solver, denoted CO-CAS, and apply it to the LiF ground-state potential energy curve \fig{fig:two_images_1}. A CAS(6,6) active space is used, and the results are compared with CASSCF.

 For the 6-31G basis set, CO-CAS is in excellent agreement with the CASSCF potential-energy curve, with energy differences within approximately 1 \si{\milli\hartree}. This agreement validates the constrained-optimization formulation in the complete-active-space limit.
 
 The behavior is different for the larger aug-cc-pVDZ basis set. In this case, the CO-CAS curve remains smooth and gives a physically reasonable equilibrium region, whereas the CASSCF calculation converges to a higher-energy solution near equilibrium. This suggests that the CASSCF iteration can be trapped at an unfavorable local stationary point for this system. In contrast, CO-CAS updates the orbital rotation through ISD on the orthogonality-constrained manifold. This behavior is consistent with previous studies of optimal-orbital selected full configuration interaction showing that direct constrained optimization of the orbital subspace can yield lower energies than conventional CASSCF-type orbital updates in some cases.

Next, we calculate the water molecule at its equilibrium geometry, with an OH bond length of 1.84345\si{\bohr} and an HOH bond angle of 110.6\si{\degree}. Using 10 active electrons, we vary the number of active orbitals and compare the ground-state energies obtained from CO-CAS and CASSCF with the cc-pVDZ and cc-pVQZ basis sets. The results are summarized in Table~\ref{tab:h2o}.

For the cc-pVDZ basis set, CO-CAS gives lower energies than CASSCF for all active spaces considered. In most cases, the energy lowering is approximately 1\si{\milli\hartree}; larger differences are found for the 16- and 18-orbital active spaces, where CO-CAS lowers the energy by about 3\si{\milli\hartree} and 2\si{\milli\hartree}, respectively. For the cc-pVQZ basis set, CO-CAS gives slightly higher energies than CASSCF for the smallest active space. As the number of active orbitals increases, however, CO-CAS again becomes energetically more favorable, yielding ground-state energies lower than those of CASSCF by approximately 1\si{\milli\hartree}.

The different behavior of CO-CAS and CASSCF can be explained by their distinct orbital-optimization formulations. In conventional CASSCF, the molecular orbitals and CI coefficients are optimized variationally within a predefined inactive-active-virtual partition. The orbital update is achieved with the rotations between different orbital subspaces that are nonredundant; rotations within the same subspace are redundant. As a result, the CASSCF orbital optimization primarily adjusts the boundaries between the inactive, active, and virtual subspaces, while the CI coefficients are optimized within the resulting CAS expansion. When the active space is compact and chemically well defined, this orbital-rotation optimization is well suited to refine the active orbitals and usually provides a robust description.

CO-CAS follows a different optimization route. For a fixed set of orbitals, the CASCI step determines the CI coefficients and the corresponding reduced density matrices (RDMs). The molecular orbitals (MOs) are then optimized directly under the orthonormality constraint using the implicit steepest descent (ISD) algorithm. In this formulation, the MO coefficients are treated as a point on the Stiefel manifold, and the ISD update follows a descent direction while preserving orbital orthonormality through projection-based constrained optimization. Thus, rather than updating only the nonredundant orbital rotations associated with a predefined subspace partition, CO-CAS directly searches for an orthonormal orbital set that lowers the CASCI energy.

This distinction becomes more important as the active space is enlarged. For small active spaces, the number of orbital degrees of freedom is limited, and the quality of the result is dominated by the initial chemical choice of the active orbitals. In this regime, the additional ISD-based orbital optimization in CO-CAS may not necessarily improve upon CASSCF, which is consistent with the slightly higher CO-CAS energies observed for the cc-pVQZ basis with 12 active orbitals. As more active orbitals are included, however, the active-virtual boundary becomes less sharply defined and the coupling between orbital optimization and static correlation becomes stronger. The orthogonality-constrained ISD optimization in CO-CAS can adapt the molecular orbitals more effectively to the correlated CASCI wave function, allowing the enlarged active space to be exploited more efficiently and give lower energies than CASSCF.

\begin{table*}[t]
	\centering
	\captionsetup{margin={1.8cm,-0cm}}
	\caption{Comparison of CO-CAS and CASSCF ground state (GS) energies for H$_2$O under the cc-pVDZ and cc-pVQZ basis set with 10 active electrons}
	\label{tab:h2o}
	\begin{threeparttable}
		
		\setlength{\tabcolsep}{14pt}

		\begin{tabular}{c c c c}
			\toprule
			& CO-CAS & CASSCF & \\
			orbs 
			& GS energy $(\si{\hartree})$ 
			& GS energy $(\si{\hartree})$ 
			& $\Delta E$ $(\si{\milli\hartree})$ \\
			\midrule
            basis set & \multicolumn{3}{c}{cc-pVDZ} \\
            \midrule
			12 & $-76.174388$ & $-76.173179$ & $-1.209$ \\
			13 & $-76.189831$ & $-76.188643$ & $-1.188$ \\
			14 & $-76.203885$ & $-76.202739$ & $-1.146$ \\
			15 & $-76.223270$ & $-76.222155$ & $-1.115$ \\
			16 & $-76.228129$ & $-76.224526$ & $-3.603$ \\
            17 & $-76.230603$ & $-76.229384$ & $-1.219$ \\
            18 & $-76.233653$ & $-76.231255$ & $-2.398$ \\
			\midrule
            basis set & \multicolumn{3}{c}{cc-pVQZ} \\
            \midrule
			12 & $-76.230799$ & $-76.235160$ & $4.361$ \\
			13 & $-76.237225$ & $-76.235698$ & $-1.527$ \\
			14 & $-76.257789$ & $-76.256436$ & $-1.353$ \\
			\bottomrule
		\end{tabular}

	\end{threeparttable}
\end{table*}

\begin{figure}[t]
		\centering
		\begin{subfigure}[b]{0.45\textwidth}
			\centering
			\includegraphics[width=\textwidth]{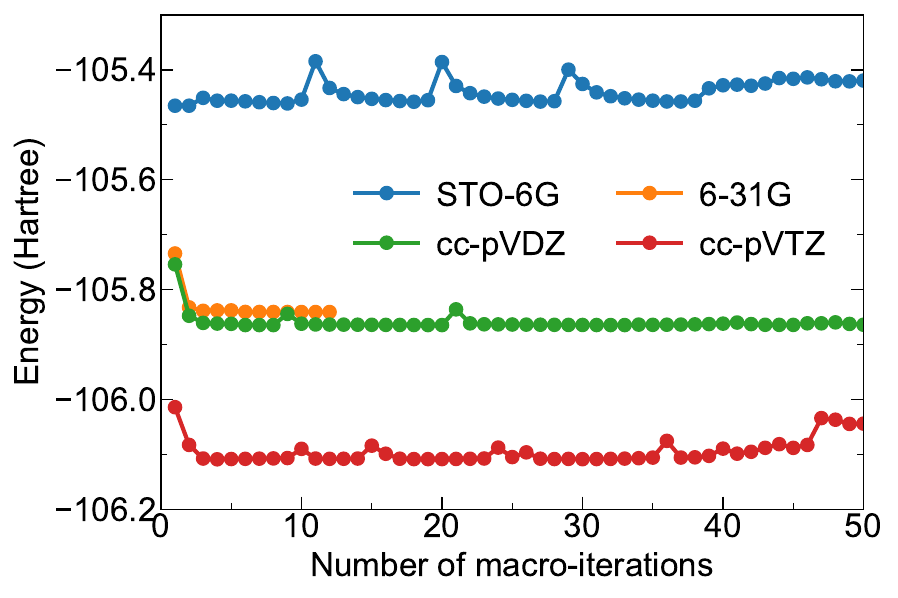}
			\caption{Without DIIS}
			\label{fig:conv_without_diis}
		\end{subfigure}
		\hfill 
		\begin{subfigure}[b]{0.45\textwidth}
			\centering
			\includegraphics[width=\textwidth]{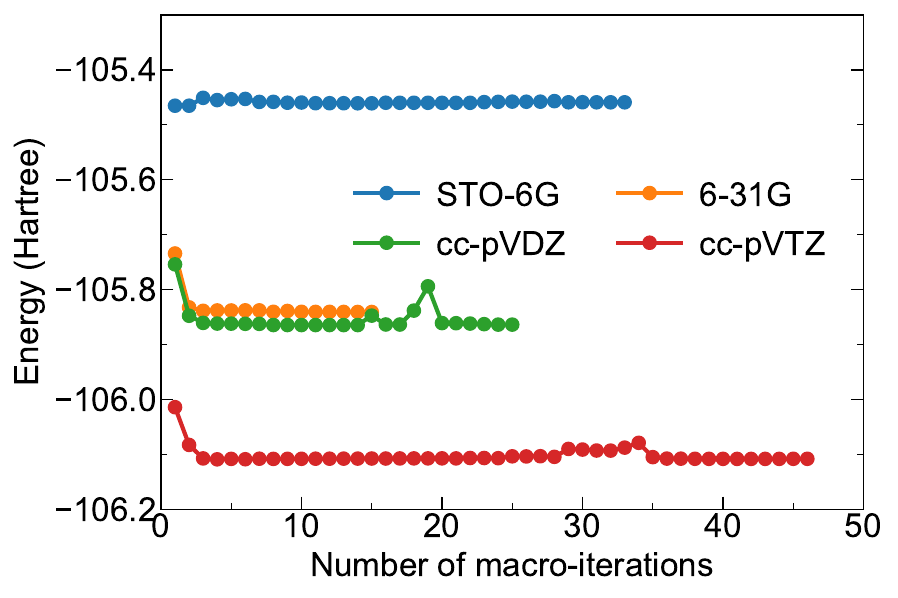}
			\caption{With DIIS}
			\label{fig:conv_with_diis}
		\end{subfigure}
		
		\caption{Convergence of LiF CO-CAS ground-state energy calculations with and without DIIS for different basis sets. The maximum number of macro-iterations is 200.}
		\label{fig:two_images_2}
	\end{figure}

To assess convergence, we compare LiF CO-CAS ground-state energy calculations at a bond length of 1.5 \si{\bohr} with and without DIIS, as shown in \fig{fig:two_images_2}. The maximum number of macro-iterations is set to 200. DIIS greatly accelerates CO-CAS convergence. Without DIIS, only the 6-31G calculation converges within the maximum number of macro-iterations. With DIIS, all tested basis sets converge within 50 macro-iterations. The improvement is significant because each macro-iteration is dominated by solving the electronic-structure problem and constructing the RDMs. Reducing the number of macro-iterations therefore directly improves the practical efficiency of the method.

	\begin{figure}[t]
		\centering
		\includegraphics[width=0.5\textwidth]{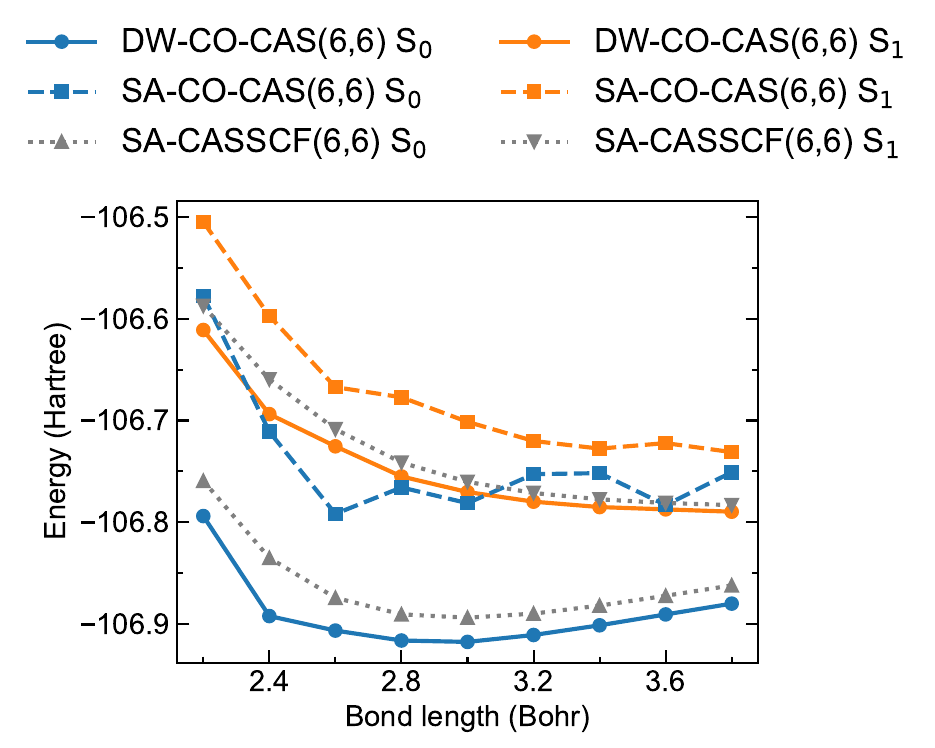} 
		\caption{LiF ground- and first-excited-state PECs calculated with different methods using the 6-31G basis set and a CAS(6,6) subspace. DW-CO-CAS denotes dynamically weighted constrained optimization configuration interaction, SA-CO-CAS denotes state-averaged constrained optimization configuration interaction, and SA-CASSCF denotes state-averaged complete active space self-consistent field.}
		\label{fig:dw}
	\end{figure}
We then examine the excited-state performance of CO-CAS(6,6)/6-31G for LiF. \fig{fig:dw} compares the ground and first excited states obtained from state-averaged CO-CAS (SA-CO-CAS), dynamically weighted CO-CAS (DW-CO-CAS), and state-averaged CASSCF (SA-CASSCF). Although state averaging is widely used for excited-state orbital optimization, the SA-CO-CAS PECs show noticeable fluctuations along the bond-stretching coordinate. 
 The DW-CO-CAS curves are smoother and more closely follow the shape of the SA-CASSCF reference. At the same time, DW-CO-CAS yields lower energies than SA-CASSCF for both states: the ground-state energy is lowered by approximately $10$ \si{\milli \hartree}, while the first excited state is lowered by approximately $1$ \si{\milli \hartree}. These results indicate that dynamical weighting provides a more flexible and stable orbital-optimization strategy for CO-CAS.

\subsection{CO-DMRG}
\begin{figure}[htbp]
		\centering
		\includegraphics[width=0.45\textwidth]{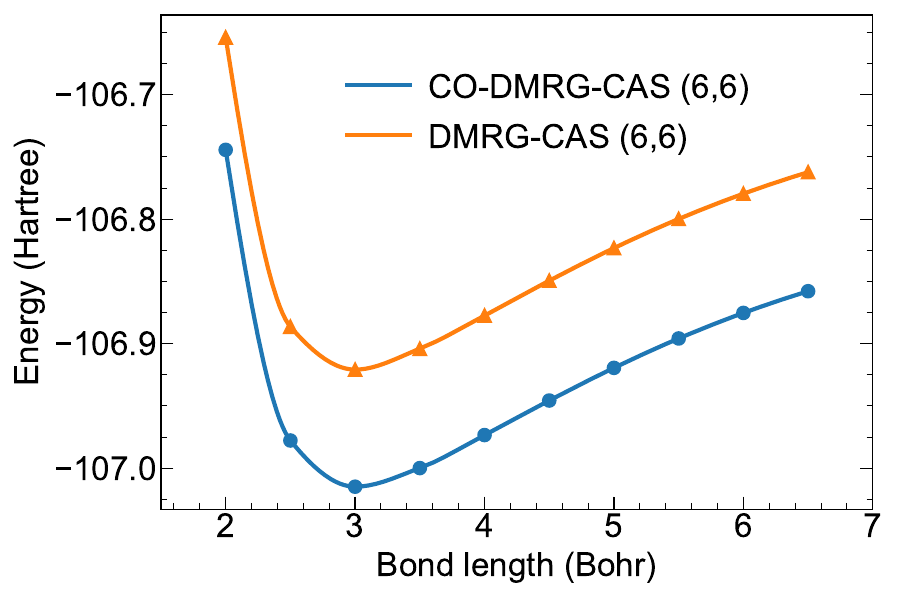} 
		\caption{LiF ground-state energies as a function of bond length calculated with CO-DMRG and DMRG in a CAS(6,6) active space with the 6-31G basis set.}
		\label{fig:fig3}
	\end{figure}

	Finally, we couple the constrained orbital optimization algorithm to DMRG and compare CO-DMRG with DMRG calculations performed without orbital optimization. The first test is the LiF ground-state PEC using a CAS(6,6) active space and the 6-31G basis set, shown in \fig{fig:fig3}.
    	At all geometries, the two curves retain similar overall shapes, but CO-DMRG yields lower energies than the DMRG calculation. The energy lowering is approximately 90 \si{\milli \hartree} over most of the curve, except at a bond length of 2 \si{\bohr}, where the lowering is approximately 30 \si{\milli \hartree}. 
 
 	\begin{figure}[t]
	 	\centering
	 	\includegraphics[width=0.45\textwidth]{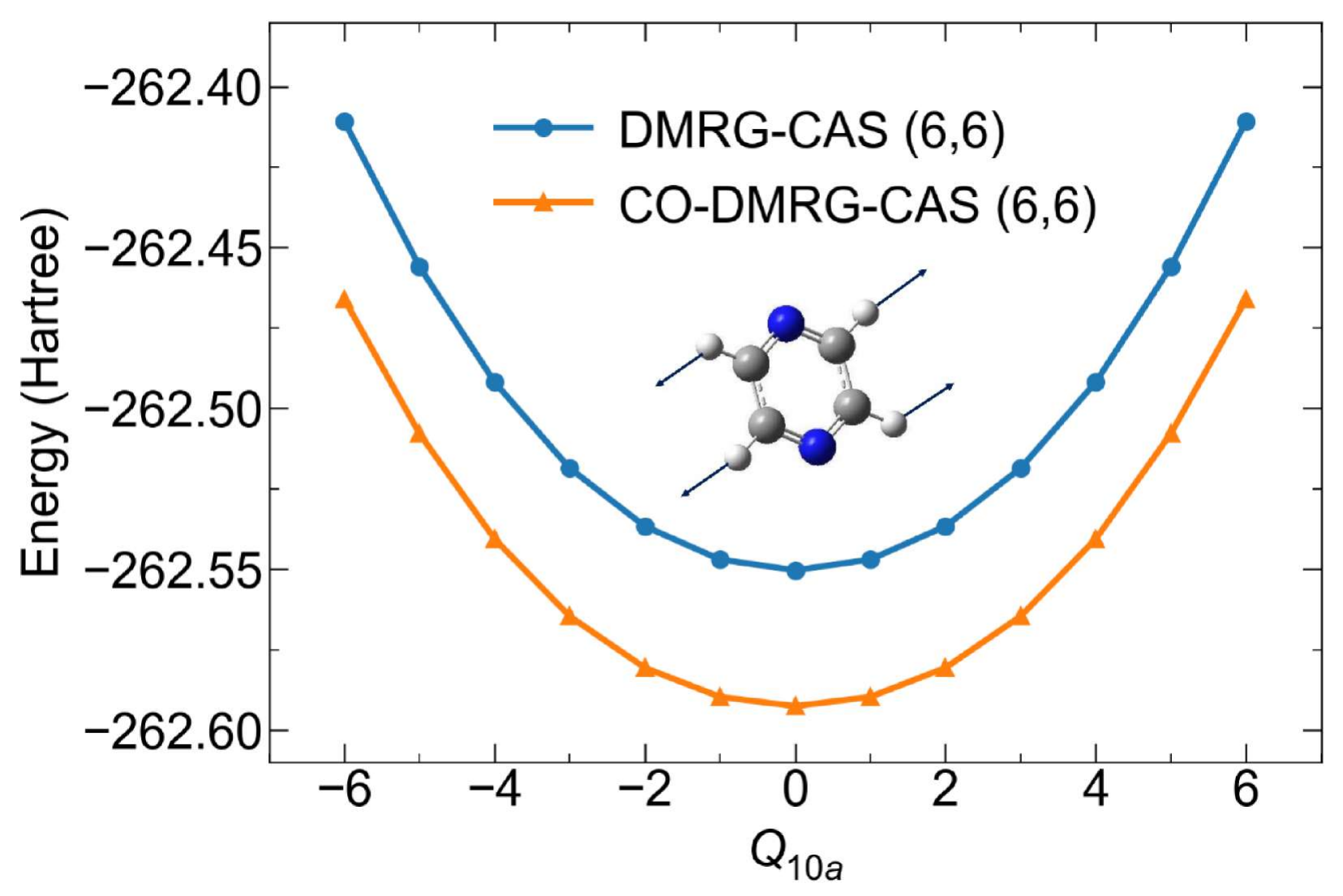} 
	 	\caption{Pyrazine potential energy curve along the $Q_{10a}$ mode calculated with CO-DMRG and DMRG in a CAS(6,6) active space with the 6-31G basis set. The white, grey, and blue spheres represent hydrogen, carbon, and nitrogen atoms, respectively. The arrows indicate the vibrational displacement vectors along the $Q_{10a}$ mode.}
	 	\label{fig:fig4}
 	\end{figure}
    	
%

	We also apply CO-DMRG to pyrazine along the $Q_{10a}$ vibrational mode \cite{shiozaki_PyrazineExcitedStates_2013}; the resulting PEC is shown in \fig{fig:fig4}. Similar to the LiF case, CO-DMRG gives lower energies than DMRG at all geometries. At $Q_{10a}=0$, the energy difference is 0.042 \si{\hartree}, and it increases with the displacement, reaching approximately 0.055 \si{\hartree} at $Q_{10a}=\pm6$. This trend indicates that orbital optimization becomes increasingly beneficial as the molecular geometry moves away from the reference structure. The optimized orbitals adapt to the changing electronic structure, allowing the fixed-active-space DMRG wavefunction to recover additional correlation energy.

	\begin{figure}[htbp]
	\centering
	\includegraphics[width=0.5\textwidth]{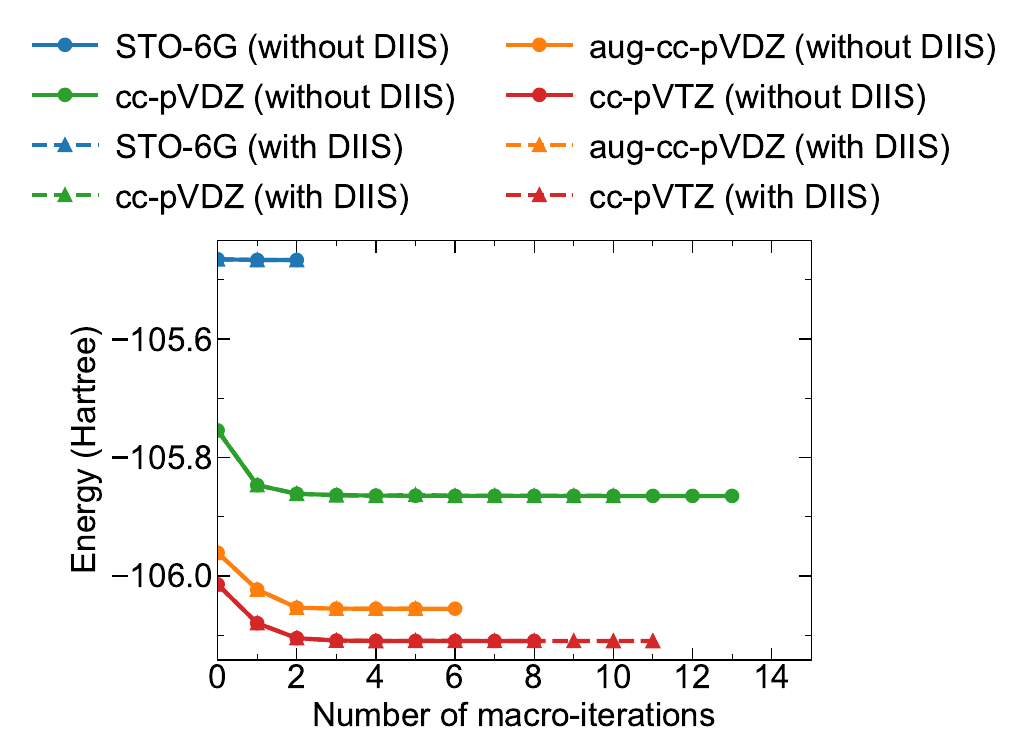} 
	\caption{Convergence of LiF CO-DMRG ground-state energy calculations with and without DIIS for different basis sets.}
	\label{fig:fig5}
	\end{figure}
	 \fig{fig:fig5} compares the convergence of CO-DMRG calculations with and without DIIS for several basis sets. In contrast to the CO-CAS case, DIIS has a smaller and less systematic effect on CO-DMRG convergence. For the STO-6G basis set, the calculation converges in two macro-iterations with or without DIIS. DIIS slightly accelerates convergence for the cc-pVDZ and aug-cc-pVDZ basis sets, but it slows convergence for cc-pVTZ in the present tests. Therefore, DIIS is clearly beneficial for CO-CAS but should be used more cautiously for CO-DMRG, where the convergence behavior also depends on the DMRG solver accuracy, bond dimension, and sweep convergence.

	\section{Conclusions}
	
We have presented a modular constrained-orbital-optimization framework for orbital optimization in quantum chemistry. The optimization is formulated as an alternating procedure consisting of two subproblems. The first subproblem solves the electronic structure problem in the current molecular orbitals, for which different solvers, including MP2, CASCI, and DMRG, can be employed. The second subproblem optimizes the molecular orbitals under the orthonormality constraint using the implicit steepest descent algorithm. This formulation separates the electronic structure solver from the orbital optimization and therefore provides a flexible interface for combining orbital optimization with different correlated wave-function methods.

Numerically, we applied the framework to ground- and excited-state potential energy curves of LiF and to ground-state calculations of pyrazine and H$_2$O. For LiF, CO-MP2 gives energies lower than conventional MP2, especially in the bond-stretching region where correlation becomes more important. CO-CAS gives results comparable to CASSCF with the 6-31G basis set and provides a smoother, lower-energy potential energy curve than the CASSCF solution obtained with the aug-cc-pVDZ basis set over the tested geometry range. For excited states, the dynamical-weighting CO-CAS scheme gives potential energy curves consistent in shape with state-averaged CASSCF while yielding lower energies. We also found that the modified DIIS procedure substantially accelerates the convergence of CO-CAS macro-iterations. For H$_2$O, CO-CAS gives lower energies than CASSCF with the cc-pVDZ basis set for all tested active spaces, and with the cc-pVQZ basis set it also yields lower energies when sufficiently large active spaces are used. In the DMRG calculations, CO-DMRG consistently lowers the energies relative to DMRG calculations performed in the initial Hartree--Fock orbital basis for both LiF and pyrazine, demonstrating that orbital optimization can improve the compactness and energetic quality of fixed-subspace tensor-network wavefunctions.

These results demonstrate that ISD-based constrained orbital optimization is a useful and general strategy for correlated electronic-structure methods. The method is not restricted to a specific wavefunction ansatz; rather, it can be combined with different electronic solvers as long as the required reduced density matrices are available. In this sense, the present formulation provides a unified framework for improving finite-subspace electronic-structure calculations under a fixed computational budget. Future work will focus on improving excited-state weighting strategies and extending the approach to photochemical and photophysical processes involving strong static correlation.

\section*{Acknowledgements}
We thank Dr. William James Glover for pointing out the dynamic weighting.

\section*{Funding}
This work is supported by the National Natural Science
Foundation of China (Grant No. 22473090 and 92356310).

\section*{Supporting Information}
Detailed derivations of the MP2, CASCI, and DMRG reduced density matrices are provided in the Supporting Information.

\section*{Data and Software Availability}

The data underlying this study are available in the Supporting Information as a compressed archive file, \texttt{data.zip}. The scripts and code used to generate and analyze the results are available on \href{https://github.com/binggu56/pyqed/tree/bg}{GitHub}.

\bibliography{reference}

\end{document}